\documentclass[11pt,twoside]{article}
\usepackage{asp2004}
\usepackage{psfig}
\usepackage{epsf}
\usepackage{graphics}
\usepackage{lscape}
\def\edcomment#1{\iffalse\marginpar{\raggedright\sl#1\/}\else\relax\fi}
\reversemarginpar

\input psfig.tex

\def\ltsima{$\; \buildrel < \over \sim \;$}
\def\simlt{\lower.5ex\hbox{\ltsima}}
\def\gtsima{$\; \buildrel > \over \sim \;$}
\def\simgt{\lower.5ex\hbox{\gtsima}}

\begin{document}
\title{Supernovae and Their Massive Star Progenitors}
\author{Alexei V. Filippenko}
\affil{Department of Astronomy, University of California, Berkeley, CA
94720-3411}

\begin{abstract} I briefly describe the Lick Observatory Supernova Search with
the 0.76-m Katzman Automatic Imaging Telescope. I then present an overview of
optical observations of Type II, IIb, Ib, and Ic supernovae (SNe), all of which
are thought to arise from core collapse in massive progenitors that have
previously experienced different amounts of mass loss. SNe~IIn are
distinguished by relatively narrow emission lines with little or no P-Cygni
absorption component; they probably have unusually dense circumstellar gas with
which the ejecta interact.  Some SNe~IIn, however, might actually be
super-outbursts of luminous variable stars; rarely, they may even be SNe~Ia in
disguise. Plausible detections of the progenitors of a few SNe~II have been
made. Spectropolarimetry of core-collapse SNe reveals that asphericity
increases toward the core.

\end{abstract}
\thispagestyle{plain}

\section{The Lick Observatory Supernova Search}

   In 1989, my team began to work on developing a robotic telescope for CCD
imaging of relatively faint objects. The history of the project is discussed in
several papers (e.g., Filippenko et al. 2001; Richmond, Treffers, \& Filippenko
1993), and several prototypes were used over the years. In 1996, we achieved
first light with our present instrument, the 0.76-m Katzman Automatic Imaging
Telescope (KAIT) at Lick Observatory on Mt. Hamilton, California. It took the
better part of another year to eliminate most of the remaining bugs in the
system, and useful scientific results started appearing in 1997. Absolutely
vital contributions to the programming and to the observing strategy were made
by Dr. Weidong Li, who joined my group in 1997.

 KAIT is a fully robotic instrument whose control system checks the weather,
opens the dome, points to the desired objects, acquires guide stars (in the
case of long exposures), exposes, stores the data, and manipulates the data
automatically, all without human intervention. We reach a limit of $\sim19$ mag
($4\sigma$) in 25-s unfiltered, unguided exposures, while 5-min guided
exposures yield $R \approx 20$ mag.  KAIT acquires well-sampled, long-term
light curves of SNe and other variable or ephemeral objects --- projects that
are difficult to conduct at other observatories having a large number of users
with different interests.

   One of our main goals is to discover nearby SNe to be used for a variety of
studies.  Special emphasis is placed on finding them well before maximum
brightness. Although the original sample of our Lick Observatory Supernova
Search (LOSS; Li et al. 2000; Filippenko et al. 2001) had only about 5000
galaxies, in the year 2000 we increased the sample to $\sim14,000$ galaxies
(most with redshift $\simlt 10,000$ km s$^{-1}$), separated into three subsets
(observing baselines of 2 days for about 100 galaxies, 3--6 days for 
$\sim3000$ galaxies, and 7--14 days for $\sim11,000$ galaxies). We are able to
observe $\sim1000$ galaxies per night in unfiltered mode. Our software
automatically subtracts new images from old ones (after registering, scaling to
account for clouds, convolving to match the point-spread-functions, etc.), and
identifies SN candidates which are subsequently examined and reported to the
Central Bureau for Astronomical Telegrams by numerous undergraduate research
assistants in my group, working with W. Li. Interested astronomers elsewhere
are also notified immediately.

   A Web page on LOSS is at
http://astro.berkeley.edu/$\sim$bait/kait.html. LOSS found its first supernova
in 1997 --- SN 1997bs; ironically, it might not even be a genuine SN (Van Dyk et al.
2000). In 1998, mostly during the second half of the year, LOSS discovered 20
SNe, thereby breaking the previous single-year record of 15 held by the Beijing
Astronomical Observatory Supernova Search.  In 1999, LOSS doubled this with 40
SNe. In 2000, LOSS found 38 SNe, even though we spent a significant fraction of
the observing time expanding the database of monitored galaxies rather than
searching for SNe. With this expanded database, LOSS discovered 68 SNe in 2001,
82 in 2002, 95 in 2003, and 71 in 2004 through October.  We discovered SN 2000A
and SN 2001A, and hence the first supernova of the new millennium, regardless
of one's definition of the turn of the millennium! During the past few years,
KAIT has discovered {\it well over half} of all nearby SNe reported world-wide,
from all searches combined. Thus, KAIT/LOSS is currently the world's most
productive search engine for nearby SNe.

 At the Lick and Keck Observatories, we spectroscopically confirm and classify
nearly all of the SNe that other observers haven't already classified. Thus,
the sample suffers from fewer biases than most.  We have started to determine
the Hubble types of the host galaxies of the SNe (van den Bergh, Li, \&
Filippenko 2002, 2003), as a first step in the calculation of rates of various
types of SNe. Already, our observations and Monte-Carlo simulations have shown
that the rate of spectroscopically peculiar SNe~Ia is considerably larger than
had previously been thought (Li et al. 2001).

Follow-up observations for the discovered SNe are emphasized during the course
of LOSS. Our goal is to build up a multicolor database for nearby
SNe. Because of the early discoveries of most LOSS SNe, our light curves
usually have good coverage from pre-maximum brightening to post-maximum
decline. Moreover, LOSS SNe are automatically monitored in
unfiltered mode as a by-product of our search; these can sometimes be useful for
other studies (e.g., Matheson et al. 2001). The positions of SNe in KAIT
images were used to identify the same SNe at very late times in {\it
Hubble Space Telescope (HST)} images (Li et al. 2002), allowing us to determine
the late-time decline rates.

   LOSS also discovers novae in nearby galaxies (e.g., M31), cataclysmic
variable stars, and occasionally comets (e.g., Li 1998). Although it records
many asteroids, we don't conduct follow-up observations, so most of them are
lost.

\section{Supernova Types}

   Supernovae occur in several spectroscopically distinct varieties; see
Filippenko (1997) for a review.  Type I SNe are defined by the absence of
obvious hydrogen in their optical spectra, except for possible contamination
from superposed H~II regions.  SNe~II prominently exhibit hydrogen in their
spectra, yet the strength and profile of the H$\alpha$ line vary widely among
these objects. At early times (within a few weeks of maximum brightness),
SNe~Ia are characterized by a deep absorption trough around 6150~\AA\ produced
by blueshifted Si~II $\lambda$6355. Members of the Ib and Ic subclasses do not
show this line. The presence of moderately strong optical He~I lines,
especially He~I $\lambda$5876, distinguishes SNe~Ib from SNe~Ic. In almost all
SNe the lines are broad, due to the high velocities of the ejecta, and most of
the lines have P-Cygni profiles formed by resonant scattering above the
photosphere.

   The late-time ($t \simgt 4$ months) optical spectra of SNe provide
additional constraints on the classification scheme. SNe~Ia show
blends of dozens of Fe emission lines, mixed with some Co lines.  SNe~Ib and
Ic, on the other hand, have relatively unblended emission lines of
intermediate-mass elements such as O and Ca. At this phase, SNe~II are
dominated by the strong H$\alpha$ emission line; in other respects, most of
them spectroscopically resemble SNe~Ib and Ic, but with narrower emission
lines. The late-time spectra of SNe~II show substantial heterogeneity, as do
the early-time spectra.

  To a first approximation, the light curves of SNe~I are all broadly similar,
although SNe~Ib usually have slower decline rates than SNe~Ic. The light curves
of SNe~II exhibit much dispersion (e.g., Patat et al. 1993), but it is useful
to subdivide the majority of them into two relatively distinct subclasses
(Barbon, Ciatti, \& Rosino 1979; Doggett \& Branch 1985).  The light curves of
SNe~II-L (``linear'') generally resemble those of SNe~I, with a steep decline
after maximum brightness followed by a slower exponential tail. In contrast,
SNe~II-P (``plateau") remain within $\sim 1$ mag (in $V\!RI$) of maximum
brightness for an extended period.  The light curve of SN 1987A, albeit
atypical, was generically related to those of SNe~II-P.

    Theoretical models are generally successful at explaining the basic
observed properties of SNe (e.g., Woosley \& Weaver 1986; Arnett et al. 1989;
Wheeler \& Harkness 1990; Woosley, Langer, \& Weaver 1995; Burrows
2000). SNe~Ia are white dwarfs in binary systems undergoing mass transfer. When
the mass of the white dwarf reaches the Chandrasekhar limit, $\sim
1.4~M_\odot$, the entire star is incinerated by a thermonuclear runaway.
Controversial issues are the nature of the burning front (subsonic, supersonic,
or a mixture?), the mass of the progenitor (can it be sub-Chandra?), and the
nature of the companion (a dwarf, giant, or white dwarf?). SNe~II, on the other
hand, result from the violent collapse of an iron core, and the subsequent
ejection of surrounding layers (largely due to neutrino interactions), in stars
having initial mass $\simgt (8-10)~M_\odot$.  SNe~Ib/Ic are probably produced
by core collapse of massive stars that have lost their outer layer of H/He,
either via winds or mass transfer to a companion.

\section{Subclasses of Type II Supernovae}

  Most SNe~II-P seem to have a relatively well-defined spectral development, as
shown in Figure 1a for SN 1999em (from Leonard et al. 2002). At early times the
spectrum is very blue, indicating a high color temperature ($\simgt$ 12,000~K;
Fig. 1b), and in some cases nearly featureless.  The temperature rapidly
decreases with time due to the adiabatic expansion and associated cooling of
the ejecta, reaching $\sim 7000$~K after a few weeks (Fig. 1b). It then
decreases more slowly during the plateau (the photospheric phase), while the
hydrogen recombination wave moves through the massive ($\sim 10~M_\odot$)
hydrogen ejecta and releases the energy deposited by the shock. At this stage
strong Balmer lines and Ca~II H\&K with well-developed P-Cygni profiles appear,
as do weaker lines of Fe~II, Sc~II, and other iron-group elements. The spectrum
gradually takes on a nebular appearance as the light curve drops to the
late-time tail; the continuum fades, but H$\alpha$ becomes very strong, and
prominent emission lines of [O~I], [Ca~II], and Ca~II also appear.

\medskip
\hbox{
\hskip -0.25truein
\psfig{figure=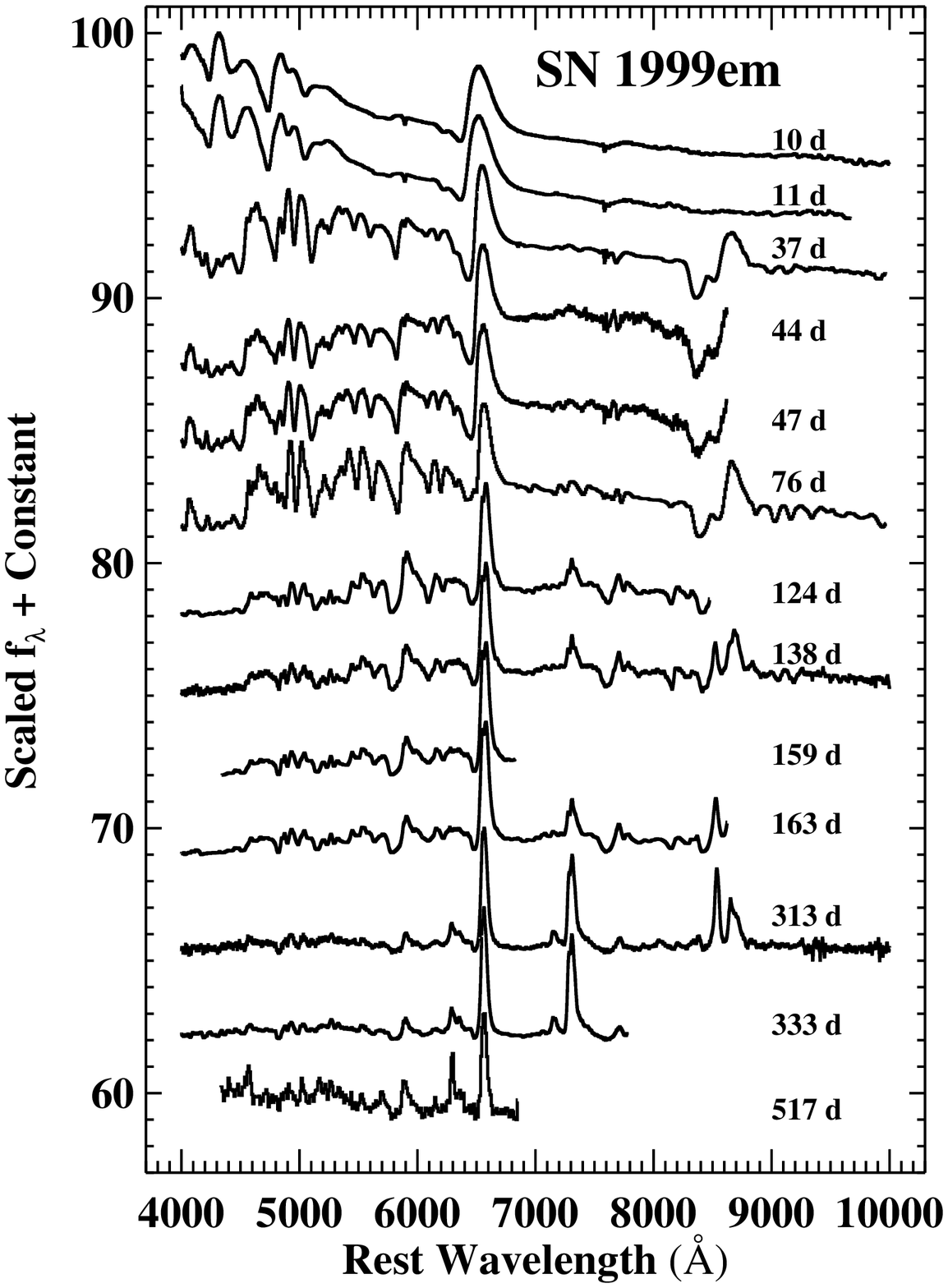,height=2.5truein,width=2.3truein,angle=0}
\hskip +0.0truein
\psfig{figure=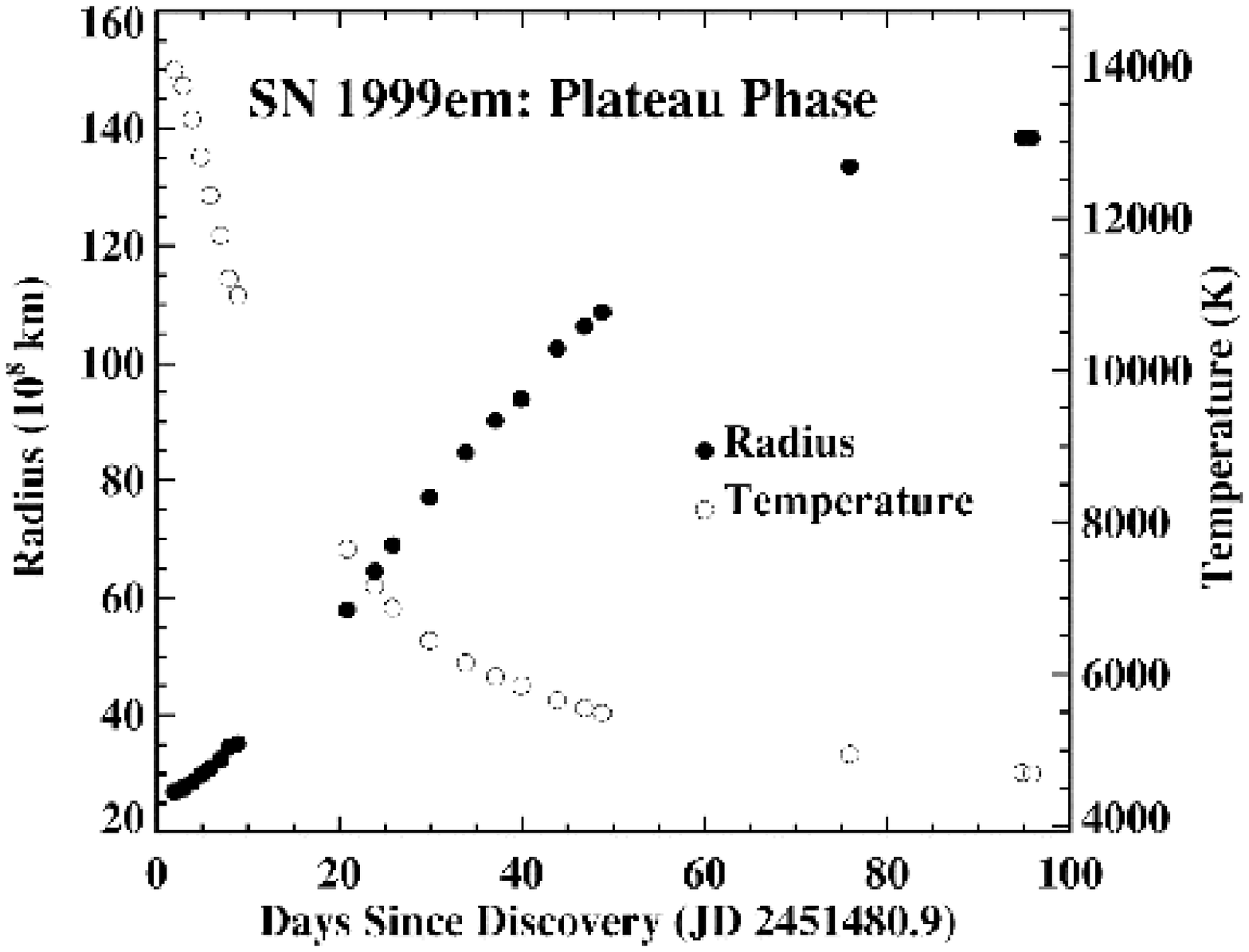,height=2.5truein,width=3.0truein,angle=0}
}

\smallskip
\noindent
{\it Figure 1:} ($a$, left) Optical spectra and
($b$, right) photospheric radius and temperature of SN II-P 1999em, from
Leonard et al. (2002).

\medskip

   Few SNe~II-L have been observed in as much detail as SNe~II-P. Branch
et al. (1981) show the spectral development of SN 1979C, an unusually luminous
member of this subclass. Near maximum brightness the spectrum is very blue and
almost featureless, with a slight hint of H$\alpha$ emission. A week later,
H$\alpha$ emission is more easily discernible, and low-contrast P-Cygni
profiles of Na~I, H$\beta$, and Fe~II have appeared. By $t \approx 1$ month,
the H$\alpha$ emission line is very strong but still devoid of an absorption
component, while the other features clearly have P-Cygni profiles. Strong,
broad H$\alpha$ emission dominates the spectrum at $t \approx 7$ months, and
[O~I] $\lambda\lambda$6300, 6364 emission is also present.  Several authors
(e.g., Wheeler \& Harkness 1990; Filippenko 1991a; Schlegel 1996)
have speculated that the absence of H$\alpha$
absorption spectroscopically differentiates SNe~II-L from SNe~II-P, but the
small size of the sample of well-observed objects precluded definitive
conclusions.

   The progenitors of SNe~II-L are generally believed to have relatively
low-mass hydrogen envelopes (a few $M_\odot$); otherwise, they would exhibit
distinct plateaus, as do SNe~II-P. On the other hand, they may have more
circumstellar gas than do SNe~II-P, and this could give rise to the
emission-line dominated spectra. Also, the light curves of some SNe~II-L reveal
an extra source of energy: after declining exponentially for several years, the
H$\alpha$ flux of SN 1980K reached a steady level, showing little if any
decline thereafter (Uomoto \& Kirshner 1986; Leibundgut et al. 1991). The
excess almost certainly comes from the kinetic energy of the ejecta being
thermalized and radiated due to an interaction with circumstellar matter
(Chevalier 1990; Leibundgut 1994).  SNe~II-L are often radio and X-ray sources,
providing further evidence for circumstellar interaction.

   During the past 15 years, there has been the gradual emergence of a new,
distinct subclass of SNe~II (e.g., Schlegel 1990; Filippenko 1991a,b;
Leibundgut 1994) whose ejecta are believed to be {\it strongly} interacting
with dense circumstellar gas, even at early times (unlike SNe~II-L). The
derived mass-loss rates for the progenitors can exceed $10^{-4} M_\odot$
yr$^{-1}$ (Chugai 1994). In these objects, the broad absorption components of
all lines are weak or absent throughout their evolution.  Instead, their
spectra are dominated by strong emission lines, most notably H$\alpha$, having
a complex but relatively narrow profile. Although the details differ among
objects (e.g., Filippenko 1997), H$\alpha$ typically exhibits a very narrow
component (FWHM $\simlt 200$ km s$^{-1}$) superposed on a base of intermediate
width (FWHM $\approx$ 1000--2000 km s$^{-1}$; sometimes a very broad component
(FWHM $\approx$ 5000--10,000 km s$^{-1}$) is also present. This subclass was
christened ``Type IIn" (Schlegel 1990), the ``n" denoting ``narrow" to
emphasize the presence of the intermediate-width or very narrow emission
components.

   In some cases, there is evidence that much of the circumstellar material of
SNe~IIn was produced quite suddenly, just a short time before the SN
explosion. For example, Chugai et al. (2004) found that the plateau-like light
curve of SN~IIn 1994W is powered by the combination of the internal energy
leakage after the explosion of an extended progenitor ($\sim 10^{15}$ cm) and
subsequent luminosity from circumstellar interaction.  The recovered
pre-explosion kinematics of the circumstellar envelope is close to homologous
expansion with an outermost velocity of $\sim 1100$ km s$^{-1}$ and a kinematic
age of $\sim 1.5$ yr.  The high mass ($\sim 0.4~M_{\odot}$) and kinetic energy
($\sim 2 \times 10^{48}$ erg) of the circumstellar envelope, combined with the
small age, strongly suggest that the circumstellar envelope was violently
ejected $\sim 1.5$ yr prior to the SN explosion.

   As briefly mentioned in Section 1, in some cases a SN~IIn is probably not a
genuine supernova (the final explosion of a star at the end of its life), but rather
an ``impostor" --- the powerful outburst of a very massive, evolved star such
as a luminous blue variable. Examples of these objects, to be discussed more fully
in the contribution by Schuyler Van Dyk, are SN 2000ch (Filippenko 2000; Wagner
et al. 2004) and SN 2002kg (Schwartz et al. 2003). This idea has been discussed
since the work of Goodrich et al. (1989) on SN 1961V.

    More recently, it has been suggested that a small subset of SNe~IIn are
actually SNe~Ia whose ejecta are interacting with dense circumstellar
material. Specifically, the early-time spectrum of SN 2002ic (Hamuy et
al. 2003) resembled that of the peculiar SN~Ia 1991T, but as the object aged
its spectrum transformed into that of a SN~IIn. A likely interpretation is that
the progenitor of SN 2002ic had an asymptotic giant branch companion that
produced a dense circumstellar environment.  Other SNe~IIn that may have been
similarly ``cloaked" SNe~Ia are SN 1997cy and SN 1999E (Wang et al. 2004).

\section{Stripped Core-Collapse Supernovae}

   Although core-collapse SNe present a wide range of spectral and photometric
properties, there is growing consensus that much of this variety is due to the
state of the progenitor star's hydrogen and helium envelopes at the time of
explosion. Those stars with massive, intact envelopes produce Type II-plateau
SNe, those that have lost their entire hydrogen envelope (perhaps through
stellar winds or mass transfer to a companion) result in SNe~Ib, and those that
have been stripped of both hydrogen and most (or all) of their helium produce
SNe~Ic; see Filippenko (1997) for a review.

   A large, comprehensive study of SNe~Ib and SNe~Ic was completed by Matheson
et al. (2001). The relative depths of the helium absorption lines in the
spectra of the SNe~Ib appear to provide a measurement of the temporal evolution
of the SN, with He~I $\lambda$5876 and He~I $\lambda$7065 growing in
strength relative to He~I $\lambda$6678 over time.  Some SNe~Ic show evidence
for weak He~I absorption, but most do not.  Aside from the presence or absence
of the helium lines, there are other spectroscopic differences between SNe~Ib
and SNe~Ic.  On average, the O~I $\lambda$7774 line is stronger in SNe~Ic than
in SNe~Ib.  In addition, the SNe~Ic have distinctly broader emission lines at
late times, indicating either a generally larger explosion energy and/or a
lower envelope mass for SNe~Ic than for SNe~Ib. These results are consistent
with the idea that the progenitors of SNe~Ic are massive stars that have lost
more of their envelope (i.e., much of the helium layer) than the progenitors of
SNe~Ib.

   The general hypothesis that SNe~Ib/Ic have ``stripped" progenitors is
greatly supported by the discovery of links between SNe~II and SNe~Ib/Ic.  For
example, as discussed by Filippenko (1988), near maximum brightness SN 1987K
was undoubtedly a SN~II, but many months later its spectrum was essentially
that of a SN~Ib. The simplest interpretation is that SN 1987K had a meager
hydrogen atmosphere at the time it exploded; it would naturally masquerade as a
SN~II for a while, and as the expanding ejecta thinned out the spectrum would
become dominated by emission from deeper and denser layers. The progenitor was
probably a star that, prior to exploding via iron core collapse, lost almost
all of its hydrogen envelope either through mass transfer onto a companion or
as a result of stellar winds. Such SNe were dubbed ``SNe~IIb" by Woosley et
al. (1987); had the progenitor of SN 1987K lost essentially {\it all} of its
hydrogen prior to exploding, it would have shown the optical characteristics of
SNe~Ib.

   The data for SN 1987K were rather sparse, making it difficult to model in
detail.  Fortunately, the Type II SN 1993J in NGC 3031 (M81) came to the rescue,
and was studied in greater detail than any supernova since SN 1987A (see
Filippenko \& Matheson 2004 for a review). Its light curves and spectra amply
supported the hypothesis that the progenitor of SN 1993J probably had a
low-mass (0.1--0.6~$M_\odot$) hydrogen envelope above a $\sim 4~M_\odot$ He
core. Filippenko, Matheson, \& Ho (1993) illustrate several early-time spectra
of SN 1993J, showing the emergence of He~I features typical of
SNe~Ib. Considerably later (Filippenko, Matheson, \& Barth 1994), the H$\alpha$
emission nearly disappeared, and the spectral resemblance to SNe~Ib was strong.
After correcting pre-explosion images for contamination from foreground stars,
Van Dyk et al. (2002) found that the energy distribution of the SN 1993J
progenitor is consistent with that of an early K-type supergiant star with $M_V
= -7.0$ mag and an initial mass of 13--22~$M_\odot$.  A star of such low mass
cannot shed nearly its entire hydrogen envelope without the assistance of a
companion star. Thus, the progenitor of SN 1993J probably lost most of its
hydrogen through mass transfer to a bound companion 3--20~AU away. Very
recently, Maund et al. (2004) reported the probable detection of hydrogen
Balmer lines from the putative companion of the progenitor of SN 1993J, thereby
providing strong evidence for the binary-star hypothesis.

  In a related study, Matheson et al. (2000) suggested a possible spectroscopic
link between SNe~Ib and SNe~Ic: the spectrum of the SN~Ic 1999cq exhibited
intermediate-width emission lines of {\it helium}, but with no corresponding
hydrogen, much as would be expected if the ejecta of the SN were interacting
with dense clumps of helium in the circumstellar medium. These clumps are
probably from the nearly pure helium layer of the massive progenitor, lost
prior to the explosion either through winds or via mass transfer to a companion
star. The remaining progenitor of the SN may have had little (if any) helium in
a shell around the C-O core. In any case, there is now little doubt that most
SNe~Ib, and probably SNe~Ic as well, result from core collapse in stripped,
massive stars, rather than from the thermonuclear runaway of white dwarfs.

\section{Evidence for CNO Processing in SN II Progenitors}

   One of the most important indicators of the extent of pre-supernova mass
loss in massive stars is the relative abundances of the CNO elements. Depending
on the mass lost and the degree of mixing, CNO burning products may be seen in
either the circumstellar medium or the outer parts of the progenitor, and
therefore in the SN. Evidence for such CNO processing has earlier been seen in
SN II-L 1979C (Fransson et al. 1984) and SN II-P 1987A (Fransson et al. 1989). More
recently, a combination of {\it HST} and ground-based
observations have revealed CNO processing in SN IIb 1993J and SN IIn 1998S
(Fransson et al. 2004), as well as in SN IIn 1995N (Fransson et al. 2002).

\vskip -0.1 truein

\begin{table}[!ht]
\caption{Summary of CNO Abundances in SNe~II}
\smallskip
\begin{center}
{\small
\begin{tabular}{lccrrl}
\tableline
\noalign{\smallskip}
Object & Type & Environment & N/C & N/O & Notes \\
\noalign{\smallskip}
\tableline
\noalign{\smallskip}
 ~SN 1979C    & II-L&ejecta&  8&$> 2$&\\
 ~SN 1987A     &II-P&circumstellar&  7.8 & 1.6&nebular analysis \\
              &   &                 &  5.0 & 1.1&photoionization model\\
 ~SN 1993J  &IIb&ejecta&12.4&$>0.8$&\\
 ~SN 1995N  &IIn&ejecta&3.8&0.2&uncertain\\
 ~SN 1998S  &IIn &circumstellar&6.0&$> 1.4$& \\
 ~Solar &  &  &  0.25 & 0.12\\
\noalign{\smallskip}
\tableline
\end{tabular}
}
\end{center}
\end{table}

\vskip -0.2 truein

 Table 1 (from Fransson et al. 2004) summarizes the derived CNO ratios for
these SNe. Because of blending, the broad-line (i.e., ejecta) determinations
are affected by a larger uncertainty (especially in the case of SN 1995N),
compared to values from the narrow circumstellar emission lines in SN 1987A and
SN 1998S.  In all cases, the N/C ratio is considerably larger than the solar
value, N/C $=0.25$ (Grevesse \& Sauval 1998). The N/O ratio is more uncertain
due to the problems with O~III] $\lambda1664$, but again it appears to be much
larger than the solar value N/O $=0.12$.

   All of the SNe in Table 1 are believed to have had progenitors that
underwent extensive mass loss prior to the explosion. Thus, it is tempting to
see the nitrogen enrichment as being a result of the mass loss.  A more
quantitative comparison, however, is not straightforward (see Fransson et al.
2004, for more details and references). Stellar evolutionary models of massive
stars, including effects of mass loss, rotation, and binarity, have been
calculated by several groups. In particular, rotation can have a large effect
on the CNO abundances by increasing the mixing from the CNO burning region
already in the main-sequence phase. Binary mass exchange produces a result
which, unfortunately, is not easily distinguished from that of rotation.

\section {The Progenitors of Core-Collapse SNe}

   Identifying the massive progenitor stars that give rise to core-collapse SNe
is one of the main pursuits of SN and stellar evolution studies.  Using
ground-based images of recent, nearby SNe obtained primarily with KAIT,
astrometry from the Two Micron All Sky Survey, and archival images from
{\it HST}, we have attempted the direct identification of
the progenitors of 16 Type II and Type Ib/c SNe (Van Dyk, Li, \& Filippenko
2003a).

   We may have identified the progenitors of the Type II SNe 1999br
in NGC 4900, 1999ev in NGC 4274, and 2001du in NGC 1365 as supergiant stars
with $M^0_V\approx -6$ mag in all three cases.  We may have also identified the
progenitors of the Type Ib SNe 2001B in IC 391 and 2001is in NGC 1961 as very
luminous supergiants with $M^0_V \approx -8$ to $-9$ mag, and possibly the
progenitor of the Type Ic SN 1999bu in NGC 3786 as a supergiant with
$M^0_V\approx -7.5$ mag.

    Additionally, we have recovered at late times SNe 1999dn in NGC 7714, 2000C
in NGC 2415, and 2000ew in NGC 3810, although none of these had detectable
progenitors on pre-supernova images.  In fact, for the remaining SNe only
limits can be placed on the absolute magnitude and color (when available) of
the progenitor.  The detected Type II progenitors and limits are consistent
with red supergiants as progenitor stars, although possibly not as red as we
had expected.  Our results for the SNe~Ib/c do not strongly constrain either
Wolf-Rayet stars or massive interacting binary systems as progenitors.

  New images of SN 2001du (in NGC 1365) available in the {\it HST} archive
allowed us to pinpoint the exact location of the SN on the pre-explosion images
(Van Dyk, Li, \& Filippenko 2003b; see also Smartt et al. 2003).  We showed
that the SN occurred in very close proximity to one of our blue candidate
stars, but we argued that this star is not the actual progenitor.  Instead, the
progenitor was not detected on the pre-SN images, and we constrained the
progenitor's initial (zero-age main sequence) mass to be less than
$13^{+7}_{-4}~M_\odot$).

  {\it HST} archival images obtained within one year prior to the explosion of
the nearby Type II-P SN 2003gd in M74 have been analyzed, and two plausible
candidates for the progenitor star were found (Van Dyk, Li, \& Filippenko
2003c). The most likely of the two progenitor candidates is a red supergiant
with initial mass of 8--9~$M_\odot$. Independently, Smartt et al. (2004)
identified the same star as the plausible progenitor of SN 2003gd. They had the
benefit of an additional, late-time {\it HST} image of SN 2003gd showing the SN
to be positionally coincident with the red supergiant. Their derived mass for
the progenitor, $8^{+4}_{-2}~M_\odot$, agrees with that of Van Dyk et
al. (2003c).  These mass estimates are somewhat lower than, but relatively
consistent with, recent limits placed on the progenitor masses of other
SNe~II-P, suggesting that such SNe arise from the iron core collapse of massive
stars at the lower extreme of the possible mass range.

\section {Spectropolarimetry of Supernovae}

    As first pointed out by Shapiro \& Sutherland (1982; see also McCall 1984),
polarimetry of a young SN is a powerful tool for probing its geometry. A hot
young SN atmosphere is dominated by electron scattering, which by its nature is
highly polarizing.  Indeed, if we could resolve such an atmosphere, we would
measure changes in both the position angle and strength of the polarization as
a function of location in the atmosphere.  For a spherical source that is
unresolved, however, the directional components of the electric vectors cancel
exactly, yielding zero net linear polarization.  If the source is aspherical,
incomplete cancellation occurs, and a net polarization results.  In general,
linear polarizations of $\sim 1$\% are expected for moderate ($\sim 20$\%) SN
asphericity.

   Filippenko \& Leonard (2004) review the results of our group's
spectropolarimetric studies of SNe. Briefly, we find that SNe~IIn tend to be
highly polarized, perhaps in part because of the interaction of the ejecta with
an asymmetric circumstellar medium. In contrast, SNe~II-P are not polarized
much, at least shortly after the explosion. At later times, however, there is
evidence for increasing polarization, as one views deeper into the expanding
ejecta.  Moreover, stripped core-collapse SNe tend to show substantial
polarization; the deeper we probe into these objects, the greater the
asphericity.  


\begin{quote}
{\bfseries Acknowledgments.} My recent research on nearby SNe has been financed by
NSF grant AST-0307894, as well as by NASA grants AR-9953, AR-10297, and
GO-10272 from the Space Telescope Science Institute, which is operated by AURA,
Inc., under NASA Contract NAS5-26555.  KAIT and its associated science have
been made possible with funding or donations from NSF, NASA, the Sylvia and Jim
Katzman Foundation, Sun Microsystems Inc., Lick Observatory, the
Hewlett-Packard Company, Photometrics Ltd., AutoScope Corporation, and the
University of California.
\end{quote}

\end{document}